\documentclass[aps,pre,twocolumn,float]{revtex4-1}

\usepackage{xcolor,hyperref}
\hypersetup{
   colorlinks,
   linkcolor={blue!50!black},
   citecolor={blue!50!black},
   urlcolor={blue!80!black}
}

\usepackage{amsmath}
\usepackage{amssymb}
\usepackage{color}
\usepackage{numprint}
\usepackage{physics}
\usepackage{hyperref}
\usepackage{bm}
\usepackage{graphicx}
\usepackage{placeins}

\npthousandsep{\,}
\DeclareMathAlphabet{\mathitbf}{OML}{cmm}{b}{it}

\newcommand{\xv}{\mathitbf x}

\newcommand{\dbar}{{\,\mathchar'26\mkern-12mu d}}
\newcommand{\Tx}{T_{\mbox{\tiny$\mathsf{X}$}}}

\begin{document}
\title{An energy-landscape-based crossover temperature in glass-forming liquids}

\author{Karina Gonz\'alez-L\'opez}
\affiliation{Institute for Theoretical Physics, University of Amsterdam, Science Park 904, Amsterdam, Netherlands}
\author{Edan Lerner}
\email{e.lerner@uva.nl}
\affiliation{Institute for Theoretical Physics, University of Amsterdam, Science Park 904, Amsterdam, Netherlands}

\begin{abstract}
The systematic identification of temperature scales in supercooled liquids that are key to understanding those liquids' underlying glass properties, and the latter's formation-history dependence, is a challenging task. Here we study the statistics of particles' squared displacements $\delta r^2$ between equilibrium liquid configurations at temperature $T$, and their underlying inherent states, using computer simulations of 11 different computer-glass-formers. We show that the relative fluctuations of $\delta r^2$ are nonmonotonic in $T$, exhibiting a maximum whose location defines the crossover temperature $\Tx$. Therefore, $\Tx$ marks the point of maximal heterogeneity during the process of tumbling down the energy landscape, starting from an equilibrium liquid state at temperature $T$, down to its underlying inherent state. We extract $\Tx$ for the 11 employed computer glasses, ranging from tetrahedral glasses to packings of soft elastic spheres, and demonstrate its usefulness in putting the elastic properties of different glasses on the same footing. Interestingly, we further show that $\Tx$ marks the crossover between two distinct regimes of the mean $\langle\delta r^2\rangle$: a high temperature regime in which $\langle\delta r^2\rangle$ scales approximately as $T^{0.5}$, and a deeply-supercooled regime in which $\langle\delta r^2\rangle$ scales approximately as $T^{1.3}$. Further research directions are discussed. 
\end{abstract}

\maketitle

\section{Introduction}
What are the important characteristic temperature scales of equilibrium liquids, that are relevant for glass formation and glass elasticity? This long-lasting question remains debated, as no single definition of a characteristic temperature scale is widely agreed upon. In the context of the dynamics of supercooled liquids~\cite{Cavagna_pedestrians}, the most prevalently discussed temperature scale is the Mode-Coupling Theory~\cite{gotze2008complex,Liesbeth_review,coslovich_mct_scipost_2019} (MCT) temperature $T_{\!\mbox{\tiny MCT}}$, conventionally extracted by fitting the primary relaxation time $\tau_\alpha$ of a liquid to the scaling form $\tau_\alpha\!\sim\! (T\!-\!T_{\mbox{\tiny MCT}})^{-\gamma}$~\cite{kablj2_mct_test,Flenner_mct_pre_2005} over an a priori-undetermined temperature range, where both the exponent $\gamma$ and the critical temperature $T_{\!\mbox{\tiny MCT}}$ are treated as fitting parameters. However, efforts to constrain these fits, or offer alternative schemes for extracting $T_{\!\mbox{\tiny MCT}}$, have been put forward~\cite{Liesbeth_review,coslovich_mct_scipost_2019}. 

Another important temperature scale of viscous liquids is traditionally referred to as the `onset'~\cite{Sastry1998,onset_reichman_2004,sri_entropy_onset_jcp_2017} or `crossover'~\cite{Goldstein1969,Cavagna_pedestrians,landscape_dominated_jeppe_2000} temperature $T_{\!\mbox{\scriptsize onset}}$. This temperature scale is understood to mark the crossover in the primary mechanism of supercooled liquids' structural relaxation, from non-activated at high temperatures, to activated over free energy barriers at low temperatures. The crossover at $T_{\!\mbox{\scriptsize onset}}$ has been proposed to be related to topological changes occurring between different regions of the energy landscape that are sampled at different equilibrium temperatures~\cite{Cavagna_prl_2000,Cavagna_prl_2002}.

While the primary focus amongst researchers in the field of glass physics has traditionally been devoted to discussing temperature scales extracted from and relevant to supercooled liquids' dynamics, not much attention has been directed towards understanding which temperature scales are relevant to the \emph{elastic} properties of the glassy states that underlie equilibrium configurations~\cite{DelGado_prl_2008,cge_paper,boring_paper,LB_modes_2019,pinching_pnas,sticky_spheres_part2}. In this work we propose a scheme to extract a characteristic crossover temperature scale $\Tx$ of a liquid by studying the statistics of particles' squared displacements $\delta r^2$ between equilibrium-liquid states at temperature $T$, and the underlying inherent states obtained by following steepest-descent dynamics from those equilibrium-liquid states~\cite{Sastry1998,Jund_prl_1999,tanaka_emergent_solidity_nat_comm_2020}. Those statistics reveal a rich picture regarding the relation between equilibrium states and their underlying inherent states, and its variation across different temperature regimes. Using a broad variety of computer glass formers, we show how the extracted crossover temperature scale $\Tx$ organizes elastic properties of glasses made by quenching equilibrium liquid states at parent temperature $T_p$. Finally, we discuss further interesting observations regarding the typical behavior of particles' squared displacements $\delta r^2$ between equilibrium states and their underlying glassy inherent states.

\section{models and methods}
We employ 11 different computer glass-formers, that vary between each other by their respective attraction strengths, characteristic stiffnesses, and other attributes described at length in Appendix~\ref{sec:app_models}. For each model, we created large ensembles of equilibrium configurations at various temperatures $T$, ranging from hot liquid states, to deeply supercooled viscous liquids. 

In order to measure the squared displacements $\delta r^2$ of individual particles between equilibrium configurations and their underlying inherent states~\cite{Sastry1998,Jund_prl_1999,tanaka_emergent_solidity_nat_comm_2020}, we follow steepest-descent (fully overdamped) dynamics, namely


\begin{equation}\label{foo00}
    \dot{\xv} \propto -\frac{\partial U}{\partial\xv}\,,
\end{equation}
where $\xv$ denotes particles' coordinates, and $U(\xv)$ is the potential energy. Eq.~(\ref{foo00}) is integrated forward in time by a simple Euler scheme~\cite{allen1989computer}, until the condition
\begin{equation}
    \sqrt{\frac{\frac{\partial U}{\partial\xv}\cdot\frac{\partial U}{\partial\xv}}{Nf^2}} \le 10^{-10}
\end{equation}
is met, where $f$ denotes the root-mean-square of interparticle forces. We recorded the squared displacements of at least $10^6$ particles at each temperature, for all models. 

At the end of the steepest descent dynamics, we calculate the athermal shear and bulk moduli of the resulting athermal glass-ensembles, following the formalism of~\cite{lutsko}. Different glass ensembles belonging to each of the employed models are labeled by the parent equilibrium temperature $T_p$ from which those glasses were quenched.

\begin{figure}[!h]
\centering
\includegraphics[width=1.0\linewidth]{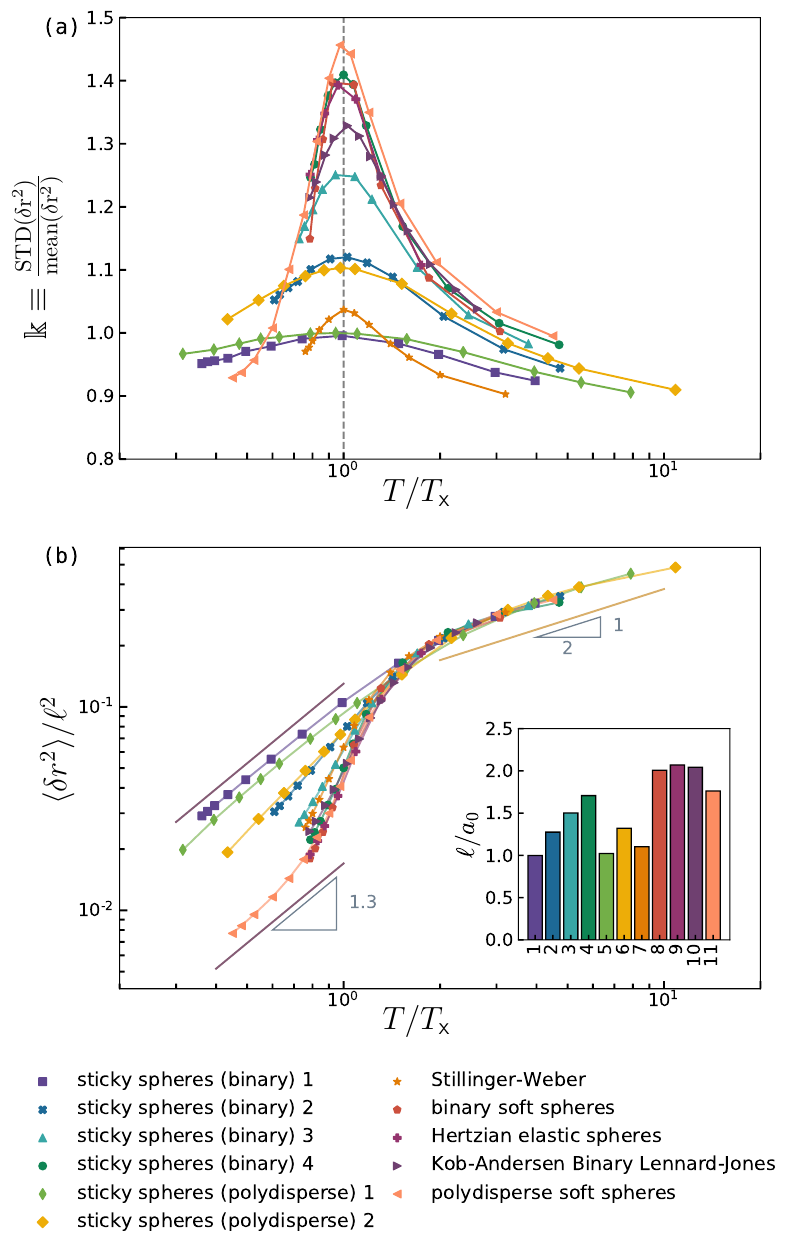}
	\caption{\footnotesize
	(a) The ratio $\Bbbk\!\equiv\!\mbox{STD}(\delta r^2)/\mbox{mean}(\delta r^2)$ quantifies the inhomogeneity of the descent process of particles from equilibrium liquid configurations at temperature $T$, to the liquids' underlying inherent states. Here we plot $\Bbbk$ vs.~$T/\Tx$ for the 11 computer-glass-forming models employed, as specified by the legend, and see Appendix~\ref{sec:app_models} for further details. The crossover temperature $\Tx$ is defined as the temperature at which $\Bbbk(T)$ assumes a maximum. (b)~The means $\langle\delta r^2\rangle$, rescaled by a (fitted, see text for details) length squared $\ell^2$ and plotted against $T/\Tx$. Inset: the length $\ell$, expressed in terms of an interparticle distance $a_0$, for all 11 glass models (see $x$-axis labels) that appear in the same order as in the legend.} 
	\label{fig:peak}
\end{figure}

\section{Relative fluctuations of $\delta r^2$ peak at $\Tx$}

Having in hand various data sets for the particle-wise squared displacements $\delta r^2$ between equilibrium states at temperature $T$, and their underlying inherent states, we next proceed to analyze the statistics of $\delta r^2$ and its temperature- and model-dependence.

In Fig.~\ref{fig:peak}a we show the ratio of the standard deviation (STD) of $\delta r^2$ to its mean, referred to in what follows as $\Bbbk(T)$. We see that $\Bbbk(T)$ is nonmonotonic for all models; it features a peak at some temperature, that we \emph{define} as $\Tx$, namely
\begin{equation}
    \frac{d\Bbbk}{dT}\bigg|_{\Tx} = 0\,.
\end{equation}
In practice we estimate $\Tx$ by fitting a cubic polynomial to the peak regime of $\Bbbk(T)$ and finding its maximum. 

The sharpness of the peak of $\Bbbk(T)$ at $\Tx$ quantified, e.g., by $\Tx^2 d^2\Bbbk/dT^2$ (evaluated at $\Tx$), constitutes an interesting dimensionless quantifier that may be useful in characterizing differences between various glass-forming models. It appears to be strongly correlated with computer glasses' thermo-mechanical annealability --- the susceptibility of glasses' mechanical properties to thermal annealing --- that was studied extensively in Ref.~\cite{sticky_spheres_part2}. It also appears to be correlated to the magnitude of the drop of the means $\langle\delta r^2\rangle$ over the $\Bbbk$-peak $T$-regime, as described next.

\section{A crossover between two regimes}

We have shown in Fig.~\ref{fig:peak}a that the relative fluctuations of $\delta r^2$ peak at $\Tx$, but how does the mean squared displacement (MSD) $\langle \delta r^2 \rangle$ behave as a function of $T$? To answer this question, we plot $\langle \delta r^2 \rangle$ vs.~$T$ in Fig.~\ref{fig:peak}b, for all models and all equilibrium temperatures $T$; the $y$-axis was rescaled by (the square of) a length $\ell$, plotted in the inset of Fig.~\ref{fig:peak}b in terms of glasses' typical interparticle distance $a_0\!\equiv\!V/N$, with $V$ and $N$ denoting the system's volume and number of particles, respectively. The lengths $\ell$ were chosen for each model such that the rescaled MSD $\langle \delta r^2 \rangle/\ell^2$ collapses at high $T$. We attribute the variability of $\ell$ to the different polydispersities used, which may hinder a direct comparison between lengths expressed in terms of an interparticle distance $a_0\!=\!(V/N)^{1/\dbar}$. 

At high temperatures, the rough collapse of the MSD shows an approximate $\sim\!T^{0.5}$ scaling. While we cannot fully explain this observation, we speculate on its relation to the Rosenfeld-Tarazona scaling~\cite{rosenfeld_1998,Jeppe_RT_scaling_2013} $\langle U \rangle\!-\! \mbox{const.}\!\sim\!T^{3/5}$: if we assume that $\langle \delta r^2 \rangle\!\sim\!\langle U \rangle\!-\!\mbox{const.}$ --- where the constant represents the energy of inherent states underlying high-$T$ equilibrium states --- then one expects $\langle \delta r^2 \rangle\!\sim\!T^{3/5}$, not far from the observed approximate scaling. Given that the energy of inherent states underlying high-$T$ equilibrium states also increases (albeit quite slowly) with $T$~(see e.g.~\cite{sri_prl_2000,onset_reichman_2004,sticky_spheres_part2}), the effective exponent characterizing the high-$T$ MSD should be smaller than 3/5, consistent with our observation and speculated picture.  

Just below the aforementioned high-$T$ scaling, the rescaled MSD $\langle \delta r^2 \rangle/\ell^2$ enters a crossover regime, that extends over a temperature range corresponding to a factor slightly larger than 2, about $\Tx$. This is a large range whose lower limit cannot be fully accessible by conventional molecular dynamics simulations (due to very large associated relaxation times), at least for a substantial subset of the glass models studied here~\cite{footnote}. The variability in accessibility to the lower limit of the crossover range indicates that the ratio $T_g/\Tx$ ($T_g$ is the glass transition temperature \cite{Cavagna_pedestrians}) is expected to vary significantly across different models, indicating that a universal relation between $\Tx$ and $T_g$ does not exist. Within the crossover regime, the rescaled MSD appears to drop by up to a factor of 10, for some models. This drop appears to be again correlated with models' thermo-mechanical annealability, shown in \cite{sticky_spheres_part2} to be controlled by the strength and form of attractive interactions in the glass. 

Finally, below the aforementioned crossover regime, some models \cite{footnote} are seen to enter the deep-supercooling regime at roughly $0.7\Tx$ in which an approximate $\langle \delta r^2 \rangle/\ell^2\!\sim\!T^{1.3}$ scaling is observed across those models. Interestingly, the prefactor of this scaling shows~a large variation across our different computer-glass-forming models, keeping in mind that rescaling $\langle \delta r^2 \rangle$ by $\ell^2$ affects possible interpretations of the prefactor variation.

\section{$\Tx$ arranges elastic properties of different glasses} \label{sec:elastic_properties}

Is the scale $\Tx$ as defined above relevant for glass physics, and if so, to which observables/phenomena? Here we show that a key dimensionless number in elasticity theory --- the ratio $G/K$ of the shear to bulk elastic moduli --- is meaningfully organized across different glass models, when plotted against $T_p/\Tx$, and rescaled by the high-$T_p$ plateau $G_\infty/K_\infty$. We reiterate that $T_p$ represents the parent equilibrium temperature from which our different glasses were quenched.  

\begin{figure}[!h]
	\centering
	\includegraphics[width=1.0\linewidth]{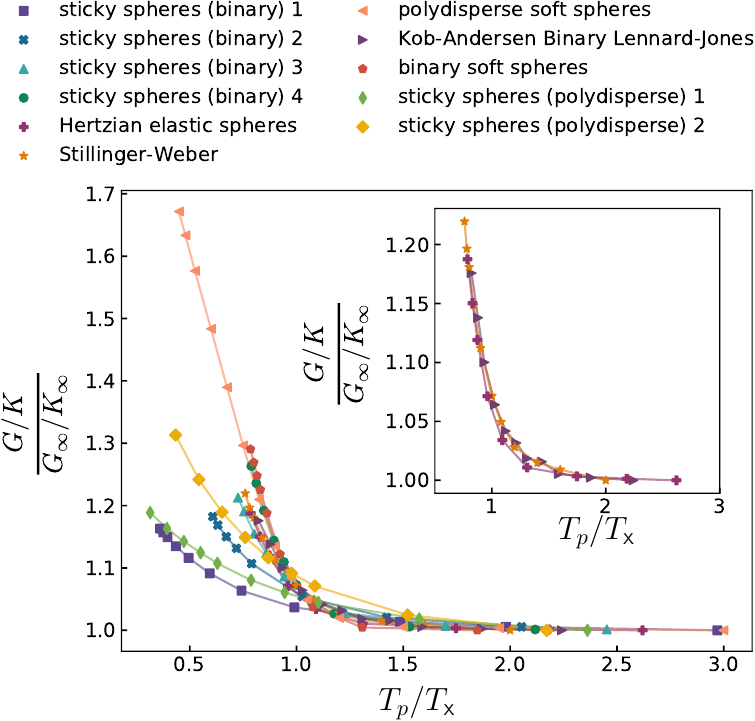}
	\caption{\footnotesize The ensemble-average shear to bulk moduli ratio $G/K$, rescaled by the high-$T_p$ plateau $G_\infty/K_\infty$, plotted against $T_p/\Tx$ for the 11 computer glass employed in our study. $\Tx$ was measured independently, as shown in Fig.~\ref{fig:peak}. Inset: same as main panel, for 3 out of the 11 glass models: elastic spheres with Hertzian interactions, the Stillinger-Weber tetrahedral glass former, and the canonical Kob-Andersen Binary Lennard-Jones model. The good collapse of $G/K$ measured for these vastly different glass models supports the utility of the crossover temperature $\Tx$ in organizing elasticity data.}
	\label{fig:G_over_K}
\end{figure}

The result is shown in Fig.~\ref{fig:G_over_K}; the scatter of the curves indicates that the rescaled $G/K$ does not generally follow a universal scaling function of $T_p/\Tx$. Nevertheless, $\Tx$ seems to represent well the crossover from the high-$T_p$ plateau to the increase at lower parents temperatures of $G/K$. In addition, one can identify subsets of models that do show a good collapse via this representation; for instance, the inset of Fig.~\ref{fig:G_over_K} shows that $\frac{G/K}{G_\infty/K_\infty}$ pertaining to soft Hertzian spheres, the Stillinger-Weber network glass former, and the Kob-Andersen Binary Lennard-Jones glass model, all collapse when plotted vs.~$T_p/\Tx$, despite the stark differences between the interaction potentials of these models (see Appendix~\ref{sec:app_models} for details). We conclude that $\Tx$ can be useful in organizing elasticity data across different models, and, in particular, plotting $\frac{G/K}{G_\infty/K_\infty}$ vs.~$T_p/\Tx$ can identify similar thermo-mechanical behaviors across different model glasses.

\section{Discussion}

In this short paper we aimed at putting forward a robust, broadly applicable and physically transparent definition of a crossover temperature $\Tx$ in supercooled liquids, which appears to be relevant to the thermal-annealing-induced variations of glasses' elastic moduli. The proposed crossover temperature $\Tx$ represents the temperature at which the relative fluctuations of the particle-wise total squared displacements $\delta r^2$, measured between equilibrium states and their underlying inherent states, is maximal. This maximum indicates that the descent process from equilibrium states to $T\!=\!0$ inherent states becomes inhomogeneous, possibly suggesting that the supercooled liquid's structure is maximally inhomogeneous too, at $\Tx$. Following the spatio-mechanical patterns that emerge during the tumbling of a system down its energy landscape might reveal the origin and essence of the observed inhomogeneities in $\delta r^2$. This interesting issue will be addressed in future work.

Apart from demonstrating the potential usefulness of the proposed crossover temperature $\Tx$, our study also reveals an interesting behavior of the means $\langle \delta r^2\rangle$, as a function of $T$ (Fig.~\ref{fig:peak}b). In particular, we show that $\Tx$ is situated in the middle of a crossover regime that separates a high-temperature regime in which $\langle \delta r^2\rangle$ scales approximately as $T^{0.5}$, and a low-temperature regime in which $\langle \delta r^2\rangle$ scales approximately as $T^{1.3}$. While we do not attribute any particular importance to the precise values of these scaling exponents, our observations nevertheless pose interesting questions regarding the nature of glass formation and its many different faces. They also suggest a possible universality in the behavior of liquids in the deep-supercooling regime $T\!\lesssim\!0.7\Tx$, which is accessible numerically via the Swap Monte Carlo algorithm~\cite{LB_swap_prx}. We note that while $\langle \delta r^2\rangle$ appears to feature 3 distinct regimes (as seen in Fig.~\ref{fig:peak}b), the ratio of $G/K$ as seen in Fig.~\ref{fig:G_over_K} does not show any clear signature of entering the deep-supercooling regime at $T_p\!\le\!0.7\Tx$. 

It is interesting to compare between our estimations for $\Tx$, and available estimations of the mode-coupling temperature $T_{\!\mbox{\tiny MCT}}$ and the onset temperature $T_{\!\mbox{\scriptsize onset}}$. For instance, the canonical Kob-Andersen Binary Lennard-Jones model~\cite{kablj} (KABLJ) --- which is also employed in our extensive computational study (see Appendix~\ref{sec:app_models}) --- has been shown to feature $T_{\!\mbox{\tiny MCT}}\!\approx\!0.435$~\cite{kablj}, in its conventional simulation units. In those same units, we have estimated $\Tx\!\approx\!0.535$, i.e.~over 20\% higher. On the other hand, in \cite{onset_reichman_2004} and in \cite{sri_entropy_onset_jcp_2017} the onset temperature for the same KABLJ model, expressed in the same simulation-units, was estimated at $T_{\!\mbox{\scriptsize onset}}\!\approx\!0.78$ and $T_{\!\mbox{\scriptsize onset}}\!\approx\!0.77$, respectively, i.e.~much higher than our estimated $\Tx$. All the aforementioned estimations of the onset or MCT temperatures, however, were not yet systematically shown to organize relaxation or elasticity data over a broad variety of glass models. Another temperature scale $T_s$ that marks the breakdown of the Stokes-Einstein relation was estimated in Ref.~\cite{Flenner_prl_2014} for the KABLJ model as $T_s\!\in\![0.55,0.60]$, which agrees reasonably well with our estimation of $\Tx$ for this model.

Finally, the observed scaling $\langle \delta r^2\rangle\!\sim\!T^{1.3}$ in the deep-supercooling regime can be very crudely interpreted as follows: $\langle \delta r^2\rangle\sim \varepsilon/\kappa$, i.e.~it is the ratio of a characteristic energy and a characteristic stiffness. In extreme supercooling conditions, i.e.~in the deep-supercooling regime, one expects the energy to follow the harmonic approximation, namely $\varepsilon\!\sim\!T$. Under these assumptions, the characteristic stiffnesses $\kappa$ is speculated to grow with decreasing temperature approximately as $\kappa\!\sim\!T^{-0.3}$. While characteristic stiffness scales are indeed observed to increase with deeper supercooling of glasses' parent equilibrium states  --- as shown in detail in Refs.~\cite{cge_paper,pinching_pnas,corrado_dipole_statistics_2020,sticky_spheres_part2} ---, the precise validation and interpretation of these speculations are left for future work.

\acknowledgments

We thank Eran Bouchbinder, Srikanth Sastry, Gustavo D\"uring, Massimo Pica Ciamarra, Geert Kapteijns, and Corrado Rainone for fruitful discussions and for their comments on the manuscript. We also warmly thank David Richard for providing data for the Stillinger-Weber model. E.~L.~acknowledges support from the NWO (Vidi grant no.~680-47-554/3259). Parts of this work were carried out on the Dutch national e-infrastructure with the support of SURF Cooperative.

\appendix
\section{Computer-glass-forming models}
\label{sec:app_models}

We employ 11 models of computer glasses in three dimensions (3D) at fixed volume $V$. In this Appendix we provide brief descriptions of the computer models employed in our work, and refer to relevant literature in which more detailed information about the models can be found. We also report the extracted value of the crossover temperature $\Tx$ for each model, expressed in terms of those models' simulation units. 

\subsection{Kob-Andersen binary Lennard Jones}
The well-known Kob-Andersen binary Lennard Jones (KABLJ) model~\cite{kablj} is a canonical glass former, which is one of the most thoroughly investigated computer glass models. It consists of an 4:1 binary mixture of type A (`large') and type B (`small') particles that interact via a radially symmetric Lennard-Jones potential. We have added a polynomial to the Lennard-Jones potential to make it smooth up to the first derivative, as done e.g.~in \cite{Leishangthem2017}. The system size employed is $N\!=\!3000$. We extracted $\Tx\!=\!0.535$ for this model.

\subsection{Stillinger-Weber model}
The Stillinger-Weber model~\cite{Stillinger_Weber} is a monocomponent liquid of $N$ particles of mass $m$ whose interaction potential consists of both a short-ranged, two-body interaction and a three-body term that favors triplets of atoms to form an angle $\theta_0\!\simeq\!109^{\mbox{\scriptsize o}}$. Further details about this model and the employed parameters can be found in \cite{modes_prl_2020}. We employed systems of $N\!=\!8000$ particles, and extracted $\Tx\!=\!0.025$ for this model.

\subsection{Binary soft spheres} \label{app:binary_ipl}
The binary soft spheres model~\cite{cge_paper} is a 50:50 binary mixture of `large' and `small' particles of equal mass $m$, interacting via an inverse-power-law (IPL) pairwise potential. Details can be found in~\cite{cge_paper}. We employed systems of $N\!=\!2000$ particles, and extracted $\Tx\!=\!0.65$ for this model. 

\subsection{Polydisperse soft spheres}
The polydisperse soft spheres model considers particles of sizes $\lambda_i$ drawn from a distribution $p(\lambda)\!\sim\!\lambda^{-3}$, between $\lambda_{\rm min}=1.0\lambdabar$ and $\lambda_{\rm max}\!=\!2.22\lambdabar$, where $\lambdabar$ is the simulation units of length. In the variant of the model we used, pairs of particles interact via the same inverse-power-law interaction potential of the Binary soft spheres model discussed above. Ensembles of equilibrium states were created using the Swap Monte Carlo Method~\cite{LB_swap_prx}, that allows for very deep supercooling. Details about the model and parameters employed here can be found in~\cite{boring_paper}. We employed systems of $N\!=\!2000$ particles, and extracted $\Tx\!=\!0.66$ for this model. 

\subsection{Hertzian spheres}
The Hertzian spheres model is a 50:50 binary mixture of small and large soft, linear-elastic spheres interacting via the Hertzian interaction law~\cite{landau1964theory}. We use systems with $N\!=\!4000$ particles, and fix the number density at $N/V\!=\!0.9386\lambdabar$, where $\lambdabar$ is the diameter of the small-particle species. We employed the same system as described in~\cite{modes_prl_2020}, and extracted $\Tx\!=\!0.0023$ for this model. 

\subsection{Sticky spheres (binary)} \label{app:ss_binary}
\begin{figure} [h!]
\includegraphics[width=0.8\linewidth]{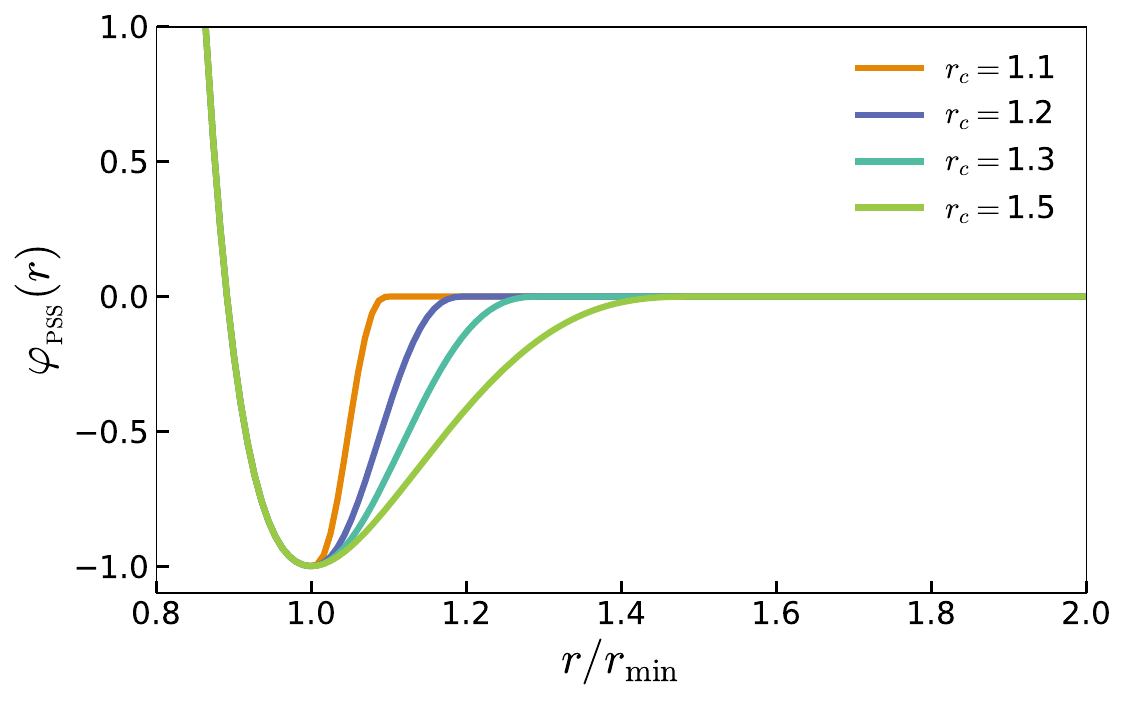}
	\caption{\footnotesize The Piecewise-Sticky-Spheres (PSS) pairwise interaction potential, used for our `sticky spheres (binary)' and `sticky spheres (polydisperse)' glass-forming models.} 
	\label{fig:ss_potential}
\end{figure}

The sticky spheres model consists of a 50:50 binary mixtures of `small' and `large' particles of equal mass $m$ in three dimensions at fixed volume $V$, interacting radially via a piece-wise modified Lennard-Jones potential --- first introduced in \cite{potential_itamar_pre_2011} --- in which the attractive part of the potential can be readily tuned via the cutoff-length $r_c$, as illustrated in Fig.~\ref{fig:ss_potential}. Further details about the interaction potential, the supercooled liquid dynamics, the elastic properties of the resulting glasses, and the precise parameters used, can be found in \cite{sticky_spheres_part2}.

Here we employ four different variants of the model, with cutoff-lengths: $r_c/r_{\rm min}\!=\!1.1, 1.2, 1.3, 1.5$ (see Fig.~\ref{fig:ss_potential} for the definition of $r_{\rm min}$), with systems of $N\!=\!3000$ particles. In the legend of Fig.1 of the main text, `sticky spheres (binary) 1,2,3,4' stand for the aforementioned model variants with $r_c\!=\!1.1, 1.2, 1.3, 1.5$, and we extracted $\Tx\!=\!2.02,1.26,1.06,0.85$, respectively, for those models. 

\subsection{Sticky spheres (polydisperse)}
This model is the same as `sticky spheres (binary)' discussed in the preceding subsection, except that instead of binary mixtures we employed the same polydispersity as used for the `polydisperse soft spheres' model discussed above. This means that particle sizes were drawn from the distribution $p(\lambda)\!\sim\!\lambda^{-3}$, between $\lambda_{\rm min}=1.0\lambdabar$ and $\lambda_{\rm max}\!=\!2.22\lambdabar$, where $\lambdabar$ is the simulation units of length. The number density $N/V$ was fixed at 0.40$\lambdabar^{-3}$, such that the high-$T_p$ pressure to bulk modulus ratio  $p/K\!\approx\!0.05$. We employed the Swap Monte Carlo algorithm~\cite{LB_swap_prx} to achieve deeply supercooled equilibrium states. The approach used to reduce possible particle-size induced finite-size effects is the same as employed in Ref.~\cite{boring_paper}. We employed two variants of this model, referred to as `sticky spheres (polydisperse) 1' using $r_c/r_{\rm min}\!=\!1.1$ and `sticky spheres (polydisperse) 2' using $r_c/r_{\rm min}\!=\!1.2$. The system size used was $N\!=\!2000$, and we extracted $\Tx\!=\!1.27$ and $\Tx\!=\!0.92$ for the $r_c/r_{\rm min}\!=\!1.1$ and $r_c/r_{\rm min}\!=\!1.2$ variants, respectively.

\clearpage
\bibliography{4-crossover_temperature_refs}

\begin{thebibliography}{36}
\bibitem [{foo()}]{footnote}%
  \BibitemOpen
  \href@noop {} {\bibfield  {journal} {\bibinfo  {journal} {We note that the
  polydisperse models employed here were equilibrated to very low temperatures
  using the Swap Monte Carlo algorithm \cite{LB_swap_prx}}\ }}\BibitemShut {NoStop}%
\end{thebibliography}

\end{document}